\documentstyle[epsbox,12pt]{article}

\begin{document}

\newcommand{\dfrac}[2]{\displaystyle{\frac{#1}{#2}}}

{\it University of Shizuoka}

\hspace*{9.5cm} {\bf US-97-02}\\[-.3in]

\hspace*{9.5cm} {\bf May 1997}\\[.3in]

\vspace*{.4in}

\begin{center}

{\large\bf  Democratic Seesaw Mass Matrix Model}\\[.1in]

{\large\bf  and New Physics
}\footnote{Talk at the Workshop on Masses and Mixings of 
Quarks and Leptons, University of Shizuoka, March 19 -- 21, 1997.
To be appeared in the Proceedings, 1997.}

\vglue.3in

{\bf Yoshio Koide}\footnote{
E-mail: koide@u-shizuoka-ken.ac.jp} \\

Department of Physics, University of Shizuoka \\ 
395 Yada, Shizuoka 422, Japan \\[.1in]

\vspace{.3in}

{\large\bf Abstract}\\[.1in]

\end{center}

\begin{quotation}
A seesaw mass matrix model is reviewed as a unification model of quark 
and lepton mass matrices.
The model can understand why top-quark mass $m_t$ is so singularly 
enhanced compared with other quark masses, especially, why $m_t \gg m_b$ 
in contrast to $m_u\sim m_d$, and why only top-quark mass is of the order 
of the electroweak scale $\Lambda_W$, i.e., $m_t\sim O(\Lambda_W)$.
The model predicts the fourth up-quark $t'$ with a mass 
$m_{t'}\sim O(m_{W_R})$, and an abnormal structure of the right-handed 
up-quark mixing matrix $U_R^u$. 
Possible new physics is discussed.
\end{quotation}

%%%%%%%%%%%%%%%%%%%%%%%%%%%%%%%%%
\newpage
%%%%%%%%%%%%%%%

\noindent{\bf 1. Why seesaw mass matrix?}
\vglue.05in

The seesaw mechanism 
$$
 M_f \simeq m_L M_F^{-1} m_R  \ .
\eqno(1.1)
$$
was first proposed [1] in order to answer the question 
why neutrino masses are so invisibly small.
And then, in order to explain why quark masses are so small compared 
with the electroweak scale $\Lambda_W$, the seesaw mechanism was 
applied  to the quarks [2]. 
However, the observation [3] of the top-quark with the large mass 
$m_t\sim O(\Lambda_W)$ brought a new situation to the seesaw mass 
matrix model: 
Why  is the top quark mass $m_t$ singularly large compared with $m_b$ 
in the third family in contrast to $m_u \sim m_d$ in the first family?
Why is the top-quark mass $m_t$ of the order of $\Lambda_W$?
It seems that the observation of the large top-quark mass rules out 
the application of the seesaw mass matrix model to  the quarks.

In the present talk, I would like to point out that the largeness 
of $m_t$, especially, $m_t\sim O(\Lambda_W)$, is rather preferable 
to the seesaw mass matrix model, and as an example, I will  review 
a specific model of a seesaw type mass matrix model, 
``democratic seesaw mass matrix model" [4,5]. 
The most of the works were done in the collaboration with 
H.~Fusaoka.
I would like to thank him for his energetic collaboration.

The basic idea is as follows.
We consider an $SU(2)_L \times SU(2)_R \times U(1)_Y$ gauge model. 
We assume vector-like fermions $F_i$ 
in addition to the three-family quarks and leptons $f_i$ 
($f=u, d, \nu, e$; $i=1,2,3$).
These fermions and Higgs scalars belong to
$$
\begin{array}{lll}
f_L = (2,1) \ , \ \ & F_L = (1,1) \ , \ \ & \phi_L=(2,1) \ ,  \\
f_R = (1,2) \ , \ \ & F_R = (1,1) \ , \ \ & \phi_R=(1,2) \ ,  \\
\end{array} \eqno(1.2)
$$
of $SU(2)_L \times SU(2)_R$. 
Then, the mass matrix for $(f, F)$ is given by
$$
M = \left(\begin{array}{cc}
0 & m_L \\
m_R & M_F \\ 
\end{array} \right) = m_0 \left( 
\begin{array}{cc}
0 & Z \\
\kappa Z & \lambda O_f \\
\end{array} \right) \ \ . 
\eqno(1.3)
$$
For simplicity, we have taken 
$$
m_L = m_R/\kappa = m_0 Z \ .
\eqno(1.4)
$$
We assume that the matrix $Z$ is universal for $f = u, d, \nu, e$.
Further, we assume that the heavy fermion mass matrix $M_F$ has 
a form [(unit matrix) + (rank-one matrix)]: 
$$
M_F = \lambda m_0 O_f = \lambda m_0 ({\bf 1} + 3 b_f  X )  \ , 
\eqno(1.5)
$$
where $b_f$ is an $f$-dependent complex parameter, 
{\bf 1} is the $3\times 3$ unit matrix, and $X$ is a rank-one matrix 
normalized by Tr$M_F=0$ at $b_f=-1/3$.
Then, for $b_f=-1/3$, we will find [4,5,6] the following mass spectrum,
$$
\begin{array}{l}
m_1, m_2 \sim \frac{\kappa}{\lambda} m_0 \ , \\
m_3 \simeq \frac{1}{\sqrt{3}} m_0 \sim O(m_L) \ , \\
m_4 \simeq \frac{1}{\sqrt{3}}\kappa m_0 \sim O(m_R) \ , \\
m_5, m_6 \sim \lambda m_0 \sim O(M_F) \ ,
\end{array} \eqno(1.6)
$$
independently of the datails of the matrix $Z$ ($\sim O(1)$).
(Also see Fig.~1 later.)
Therefore, if we assume that Tr$M_F=0$ for up-quark sector, 
we can naturally understand why only the top quark has a mass of 
the order of the electroweak scale $\Lambda_W\sim O(m_L)$.
This point will also be emphasized by T.~Satou [7] in this session
from more general study of the seesaw quark mass matrix.

%%%%%%%%%%%%%%%%%%%%%%%%%%%%%
\vglue.2in

\noindent{\bf 2. Why democratic $M_F$?}
\vglue.05in

So far, we have never assumed that
the rank-one matrix $X$ is a democratic type.
Next, I would like to talk about why our model is called 
``democratic" [8].

We know that we can always take the rank-one matrix $X$ 
as a democratic type 
$$
X = \frac{1}{3} \left(\begin{array}{ccc}
1 & 1 & 1 \\
1 & 1 & 1 \\
1 & 1 & 1 \\
\end{array} \right) \ \ , 
\eqno(2.1)
$$
without losing generality.
The naming ``democratic" for the model is motivated by the following 
phenomenological success [4] of taking $M_F$ ``democratic":
 if we assume that the matrix $Z$ is given 
by a diagonal matrix
$$
Z = \left( 
\begin{array}{ccc}
z_1 & 0 & 0 \\
0 & z_2 & 0 \\
0 & 0 & z_3 
\end{array} \right) \propto 
 \left( 
\begin{array}{ccc}
\sqrt{m_e} & 0 & \\
0 &\sqrt{m_\mu} & 0 \\
0 & 0 & \sqrt{m_\tau} 
\end{array} \right) \ . 
\eqno(2.2)
$$  
we can obtain reasonable values of the quark masses $m_i^q$ and 
Cabibbo-Kobayashi-Maskawa (CKM) [9] matrix $V$.
For example, we can obtain the successful relation [10]
$$
\frac{m_u}{m_c}\simeq \frac{3}{4}\frac{m_e}{m_\mu} \ , \eqno(2.3)
$$
for $b_u\simeq -1/3$.
So, hereafter, we call the seesaw mass matrix model 
with (2.1) and (2.2) the ``democratic seesaw mass matrix model".

Such a structure of the matrix $Z$ was suggested from 
the following phenomenology:
Experimentally well-satisfied charged lepton mass formula [11]
$$
m_\tau + m_\mu + m_e = \frac{2}{3} \left(\sqrt{m_\tau} + \sqrt{m_\mu} 
+ \sqrt{m_e} \right)^2 
\eqno(2.4)
$$
can be derived from the bi-liner form 
$$
M_e \propto Z \cdot {\bf 1} \cdot Z \ ,
\eqno(2.5)
$$
where
$$
Z = \left( 
\begin{array}{ccc}
z_1 & 0 & 0 \\
0 & z_2 & 0 \\
0 & 0 & z_3 
\end{array} \right) \ , \ \  \ \ 
\begin{array}{l}
z_i\equiv x_i +x_0  \ , \\
x_1+x_2+x_3=0 \ , \\
x_0^2=(x_1^2+x_2^2+x_3^2)/3 \ . 
\end{array} \eqno(2.6)
$$
The form (2.5) suggests a seesaw mass matrix model 
with a U(3)-family nonet Higgs boson [12].
However, in the present talk, I will skip this topic because 
I have no time sufficient to discuss it.

%%%%%%%%%%%%%%%%%%%%%%%%%%%%%%%%%%
\vglue.2in

\noindent{\bf 3. Phenomenology of $m_i^q$ ($q=u, d$) and $V$}
\vglue.05in

We take the rank-one matrix $X$ as the democratic form (2.1).
Then, the successful results for $m_i^q$ and $V$ are obtained 
from the following assumptions and inputs.

[Assumption I]: The matrix $Z$ takes a diagonal form 
$Z={\rm diag}(z_1, z_2, z_3)$, when $X$ is in a democratic basis (2.1).

[Assumption II]: The parameter $b_f$ takes $ b_e =0$, 
in the charged lepton sector.

The assumption II was put in order to fix the parameters $z_i$ as a trial.
Then, the parameters $z_i$ are given by
$$
\frac{z_1}{\sqrt{m_e}} = \frac{z_2}{\sqrt{m_\mu}} 
= \frac{z_3}{\sqrt{m_\tau}} 
= \frac{1}{\sqrt{m_e + m_\mu + m_\tau}} \ .
\eqno(3.1)
$$
In Fig.~1, we show the behavior of mass spectra $m_i^f$ 
($i=1,2,\cdots , 6$) versus the parameter $b_f$.
As seen in Fig.~1, the third fermion mass $m_3^f$ is sharply enhanced at 
$b_f=-1/3$.
Also, note that the masses $m_2^f$ and $m_3^f$ (masses $m_1^f$ and $m_2^f$) 
degenerate at $b_f=-1/2$ and $b_f=-1$, and the degeneration disappears 
for the case of arg$b_f=0$.
%%%%%%%%%%%%%%%%%%%%%%%%%%%%%%%%%%%%%%%%%%%%%%%%%%%%%Fig.1
\begin{figure}[htbp]
\begin{minipage}[tl]{10cm}
\epsfile{file=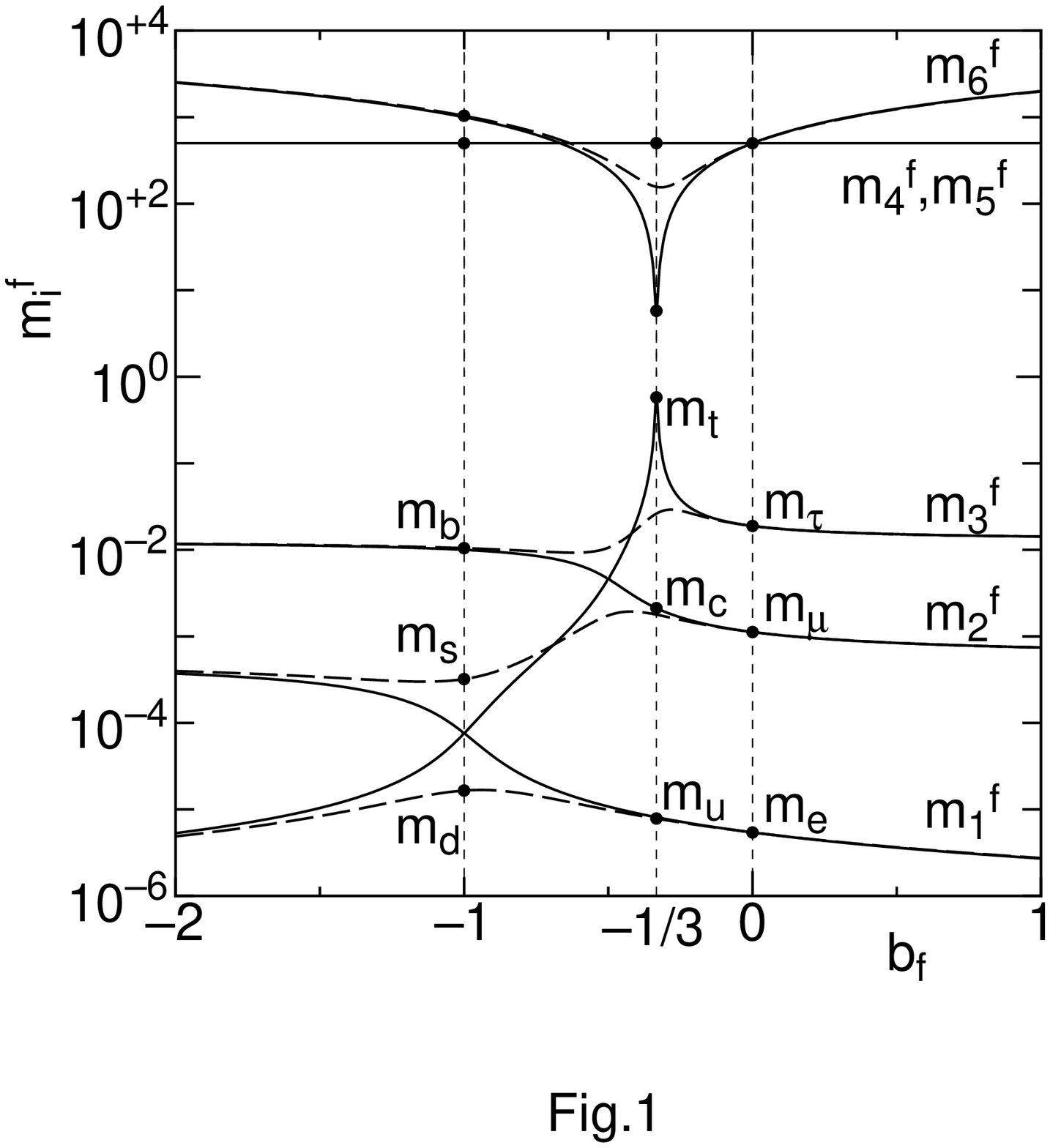,scale=0.45}
\end{minipage}
\begin{minipage}[tr]{4cm}
{\small Fig.~1. Masses $m_i$ ($i=1,2,\cdots,6$) versus $b_f$ for the 
case $\kappa=10$ and $\kappa/\lambda=0.02$.
The solid and broken lines represent the cases arg$b_f=0$ and 
arg$b_f=18^\circ$, respectively. The figure was quoted from Ref.~[5].
}
\end{minipage}
\end{figure}
%%%%%%%%%%%%%%%%%%%%%%%%%%%%%%%%%%%%%%%%%%%%%%%%%%%%%%%

In addition to (2.3), we can obtain many interesting relations [4,5]:
$$
\frac{m_c}{m_b}\simeq 4\frac{m_\mu}{m_\tau} \ , \ \ \ 
\frac{m_d m_s }{m_b^2}\simeq 4\frac{m_e m_\mu}{m_\tau^2} \ , \ \ \ 
\frac{m_u}{m_d}\simeq 3\frac{m_s}{m_c} \ , \eqno(3.2)
$$
around $b_u \sim - 1/3$, and  $b_d \sim -e^{i\beta_d}$ 
$(1\ll \beta_d^2 \neq 0)$. 
Therefore, we put the following assumption.

[Assumption III]:  We fix the values of $|b_f|$ for the quark-sector as
$$
b_u=-\frac{1}{3} \ , \ \ \ 
b_d=-e^{i\beta_d} \ (1\ll \beta_d^2 \neq 0) .
\eqno(3.3)
$$
The former means the ansatz of 
``the maximal top-quark-mass enhancement",  
but, at present, there is not good naming for the latter.

For phenomenological fitting, we have used the following inputs: 
$\kappa/\lambda=0.02$ from the observed ratio $m_c/m_t$;  
$\beta_d=18^\circ $ from the observed ratio $m_d/m_s$.
Then we obtain reasonable quark mass ratios 
and CKM matrix parameters [4]:
$$
\begin{array}{ll}
|V_{us}|=0.220 \ , \ \ \ &  |V_{cb}|=0.0598 \ , \\ 
|V_{ub}|=0.00330 \ , \ \ \ & |V_{td}|=0.0155 \ . 
\end{array} \eqno(3.4)
$$
(The value of $|V_{cb}|$ is somewhat larger than the observed value [13]
$ |V_{cb}|_{exp}=0.041\pm 0.003$. 
For the improvement of the numerical value, 
see Ref.~[5].)

%%%%%%%%%%%%%%%%%%%%%%%%%%%% 4.
\vglue.2in

\noindent{\bf 4.  Application to neutrino mass matrix}
\vglue.05in

As seen in Fig.~1, the choice of 
 $b_f\simeq -1/2$ gives 
$$
m_1 \ll m_2 \simeq m_3 \ , \eqno(4.1)
$$
$$
U_{L} \simeq \left(
\begin{array}{ccc}
1 & \frac{1}{\sqrt{2}}\left(\sqrt{\frac{m_e}{m_\mu}}- 
\sqrt{\frac{m_e}{m_\tau}}\right) 
& \frac{1}{\sqrt{2}}\left(\sqrt{\frac{m_e}{m_\mu}}+ 
\sqrt{\frac{m_e}{m_\tau}}\right)  \\
-\sqrt{m_e/m_\mu} & \frac{1}{\sqrt{2}} & 
-\frac{1}{\sqrt{2}} \\
-\sqrt{m_e/m_\tau} & \frac{1}{\sqrt{2}} & 
\frac{1}{\sqrt{2}} \\
\end{array} \right) \ .\eqno(4.2)
$$
On the other hand, the atmospheric neutrino data (Kamiokande) [14]  
have suggested a large neutrino mixing
$\sin^22\theta_{\mu\tau}\simeq 1$ with 
$\Delta m^2_{\tau\mu} \simeq 1.6 \times 10^{-2}$  ${\rm eV}^2$, and 
the solar neutrino data (with MSW effects) [15]  have suggested 
a neutrino mixing $\sin^22\theta_{e x}\simeq 0.007$ with 
$\Delta m^2_{x e} \simeq 6 \times 10^{-6}$ eV$^2$.
The results (4.1) and (4.2) are preferable to these data.

In order to make the model more explicit, we put the following assumption:
We assume that $\nu_R$ has a Majorana mass of the order of $\xi m_0$ 
($\xi\gg\lambda\gg\kappa\gg 1$) 
in addition to the heavy neutrino masses $M_N \sim O(\lambda m_0)$.
Then, for example, for $b_\nu = -0.41$,  we obtain
$$
\begin{array}{ll}
\sin^22\theta_{\mu\tau}\simeq 0.58 \ , \ \ \ \ \ &
\Delta m^2_{\tau\mu} \simeq 1.0 \times 10^{-2} \ {\rm eV}^2 \ , \\
\sin^22\theta_{e\mu}\simeq 0.0061 \ , \ \ \ \ \ &
\Delta m^2_{\mu e} \simeq 6 \times 10^{-6} \ {\rm eV}^2 \ , 
\end{array} \eqno(4.3)
$$
with $\xi m_0 = 1.9 \times 10^9$ GeV.
More details have been given in  Ref.~[16].

%%%%%%%%%%%%%%%%%%%%%%%%%%%% 5.

\vglue.2in
\noindent{\bf 5. Abnormal Structure of $U_R^u$}
\vglue.05in

In the down-quark sector,  where the seesaw expression 
$M_f \simeq m_L M_F^{-1} m_R$ is valid, 
the mixing matrices $U_L^d$ and $U_R^d$ are given by
$$ 
U_L^d = \left(
\begin{array}{cc}
A_d & \frac{1}{\lambda} C_d \\
\frac{1}{\lambda} C'_d & B_d 
\end{array} \right) \ , \ \ 
U_R^d \simeq \left(
\begin{array}{cc}
A_d^* & \frac{\kappa}{\lambda} C_d^* \\
\frac{\kappa}{\lambda} C^{\prime \ast}_d & B_d 
\end{array} \right) \ . 
\eqno(5.1)
$$
On the contrary, in the up-quark sector, 
where the seesaw expression is not valid any longer, 
the mixing matrices $U_L^u$ and $U_R^u$  are given by
$$
U_L^u = \left(
\begin{array}{lll|lll} 
+0.9994 & -0.0349 & -0.0084 & -0.0247 \frac{1}{\lambda} & 
+6\times 10^{-5}\frac{1}{\lambda} & +4\times 10^{-6}\frac{1}{\lambda} \\
+0.0319 & +0.9709 & -0.2373 & -0.2051\frac{1}{\lambda} 
& -0.4346\frac{1}{\lambda} & +0.0259\frac{1}{\lambda} \\
+0.0165 & +0.2369 & +0.9714 & +0.8990\frac{1}{\lambda} & 
+0.8431\frac{1}{\lambda} & -0.0444\frac{1}{\lambda} \\ 
\hline
+0.0934\frac{1}{\lambda} & +0.1114\frac{1}{\lambda} 
& -1.0365\frac{1}{\lambda} & +0.5774 & +0.5774 & +0.5772 \\
-0.0118\frac{1}{\lambda} & +0.1649\frac{1}{\lambda} 
& +0.0209\frac{1}{\lambda} & -0.7176 & +0.6961 & +0.0215 \\
-0.0064\frac{1}{\lambda} & -0.1011\frac{1}{\lambda} 
& +0.7927\frac{1}{\lambda} & -0.3894 & -0.4267 & +0.8163 
\end{array} \right) \ ,
\eqno(5.2)
$$
$$
U_R^u = \left(
\begin{array}{lll|lll} 
+0.9994 & -0.0349 & -0.0084 & -0.0247\frac{\kappa}{\lambda} & 
+6\times 10^{-5}\frac{\kappa}{\lambda} 
& +4\times 10^{-6}\frac{\kappa}{\lambda} \\
+0.0319 & +0.9709 & -0.2373 & -0.2051\frac{\kappa}{\lambda} 
& -0.4346\frac{\kappa}{\lambda} & +0.0259\frac{\kappa}{\lambda} \\
+0.0256\frac{\kappa}{\lambda} & +0.3459\frac{\kappa}{\lambda} 
& -0.0747\frac{\kappa}{\lambda} & +0.5773 & +0.5773 & +0.5774 \\
\hline
+0.0165 & +0.2369 & +0.9713 & +0.3274\frac{\kappa}{\lambda} 
& +0.2716\frac{\kappa}{\lambda} & -0.6160\frac{\kappa}{\lambda} \\ 
-0.0118\frac{\kappa}{\lambda} & +0.1649\frac{\kappa}{\lambda} 
& +0.0209\frac{\kappa}{\lambda} & -0.7176 & +0.6961 & +0.0215 \\
-0.0064\frac{\kappa}{\lambda} & -0.1011\frac{\kappa}{\lambda} 
& +0.7929\frac{\kappa}{\lambda} & -0.3894 & -0.4267 & +0.8161 
\end{array} \right) \ .
\eqno(5.3)
$$
Note that the right-handed up-quark mixing 
matrix $U_R^u$ has a peculiar structure 
as if third and fourth rows of $U_R^u$ are exchanged 
in contrast to $U_L^u$.

Why does such an abnormal structure appear in $U_R^u$?
In order to see this, let us change the heavy fermion basis 
from the ``democratic basis" to the ``diagonal basis":
$$
M_F\ \  \Longrightarrow \ \ \widetilde{M}_F\equiv 
A M_F A^{-1} =  m_0 \lambda \left(
\begin{array}{ccc}
1+3b_f & 0 & 0 \\
0 & 1 & 0 \\
0 & 0 & 1 
\end{array} \right) \ . 
\eqno(5.4)
$$
Then, the Hermitian matrices $H_L\equiv \widetilde{M}\widetilde{M}^\dagger$ 
and $H_R \equiv \widetilde{M}^\dagger\widetilde{M}$ take the following forms:
$$
\begin{array}{ll}
H_L = \widetilde{M}\widetilde{M}^\dagger  &
H_R = \widetilde{M}^\dagger\widetilde{M} \\
= 
m_0^2 \left(\begin{array}{cc}
\widetilde{Z}^T \widetilde{Z} & \lambda\widetilde{Z}^T\widetilde{O}_u \\ 
\lambda\widetilde{O}_u\widetilde{Z} & 
\lambda^2\widetilde{O}_u^2 + \kappa^2 \widetilde{Z}\widetilde{Z}^T \\ 
\end{array} \right)  \ & 
= m_0^2 
\left(\begin{array}{cc}
\kappa^2\widetilde{Z}^T\widetilde{Z} 
& \kappa \lambda \widetilde{Z}^T \widetilde{O}_u \\ 
\kappa\lambda \widetilde{O}_u\widetilde{Z} & 
\lambda^2 \widetilde{O}_u^2 + \widetilde{Z}\widetilde{Z}^T \\ 
\end{array} \right)   \ \\
\end{array}$$
$$
\begin{array}{ll}
=\left(
\begin{array}{ccc|ccc}
\ast & * & * & 0 & * \lambda & * \lambda \\
\ast & * & * & 0 & * \lambda & * \lambda \\
\ast & * & * & 0 & * \lambda & * \lambda  \\
\hline
0 & 0 & 0 & * \kappa^2 & * \kappa^2 & * \kappa^2 \\
\ast  \lambda & * \lambda & * \lambda & 
* \kappa^2 & * \lambda^2 & * \kappa^2 \\
\ast  \lambda & * \lambda & * \lambda & 
* \kappa^2 & * \kappa & * \lambda^2  \\
\end{array} \right) 
& 
=\left(
\begin{array}{ccc|ccc}
\ast \kappa^2 & * \kappa^2 & * \kappa^2 & 
0 & * \kappa\lambda & * \kappa\lambda  \\
\ast \kappa^2 & * \kappa^2 & * \kappa^2 & 
0 & * \kappa\lambda & * \kappa\lambda  \\
\ast \kappa^2 & * \kappa^2 & * \kappa^2 & 
0 & * \kappa\lambda & * \kappa\lambda  \\
\hline
0 & 0 & 0 & *  & *  & *  \\
\ast \kappa \lambda & * \kappa\lambda & * \kappa\lambda & 
*  & * \lambda^2 & *  \\
\ast \kappa \lambda & * \kappa\lambda & * \kappa\lambda & 
*  & * & * \lambda^2 \\
\end{array} \right) 
\end{array}
\eqno(5.5)
$$
where $\widetilde{Z}=AZ$ and $\ast \sim O(1)$.
The result(5.5) means that the top quark $t\equiv u_3$ and 
the fourth up-quark $t'\equiv u_4$  consist of 
the following components
$$
\begin{array}{l}
t \equiv u_3 \simeq (u_{3L}, U_{1R}) \ , \\
t'\equiv u_4 \simeq (U_{1L}, u_{3R}) \ , 
\end{array} \eqno(5.6)
$$
although $u\simeq (u_{1L}, u_{1R})$,  $c\simeq (u_{2L}, u_{2R})$, 
$u_5\simeq (U_{2L}, U_{2R})$ and $u_6\simeq (U_{3L}, U_{3R})$.
Therefore, we can expect a single $t'$ production 
through the exchange of the right-handed weak boson $W_R$ 
as we state later.

%%%%%%%%%%%%%%%%%%%%%%%%%%%%%%%%%%%%%%%%% 6. 
\vglue.2in
\noindent{\bf 6. New physics from DSMM}
\vglue.05in

Since we want to observe new effects from the present model, 
we take $\kappa=10$ tentatively.
Then, we can expect $m_{t'} \simeq \kappa m_t \sim $ a few TeV.
The single $t'$ production may be observed 
through the exchange of $W_R$ as $ d+u \rightarrow t' +d$, 
with $|V_{t'd}^R|= 0.0206$ and $|V_{ud}^R|=0.976$.
For example, we will observe the production
$ p +p \rightarrow t' +X$ at LHC.

On the other hand, in the present model, FCNC effects appear
proportionally to the factor [17]
$$
\xi^f = U_{fF} U_{fF}^\dagger \ , 
\ \ 
{\rm where}
\ \ 
U=\left( 
\begin{array}{cc}
U_{ff} & U_{fF} \\
U_{Ff} & U_{FF} 
\end{array} \right) \ . \eqno(6.1)
$$
Note that the FCNC effects appear visibly 
in the modes related to top-quark, because the large elements are 
only $(\xi_R^u)_{tc}=-0.00709$ and $(\xi_R^u)_{tu}=-0.000284$, and 
the other elements are harmlessly small, e.g., 
$(\xi_R^u)_{cu}=\kappa^2 (\xi_L^u)_{cu}=2.01 \times 10^{-6}$, 
$|(\xi_R^d)_{ds}|=\kappa^2 |(\xi_L^d)_{ds}|=4.03 \times 10^{-8}$, 
and so on.
For example, we may expect the single top-quark production
$ e^- + p\rightarrow e^- +t +X $ at HERA.
Unfortunately, the values $(\xi_L^u)_{tu}=-8.85 \times 10^{-8}$ and 
$(\xi_R^u)_{tu}=-2.84 \times 10^{-4}$ lead to an invisibly small value 
of the cross section $\sigma (e^-+ p\rightarrow e^- + t+ X) \sim 10^{-8}$ pb.
Only possibility of the observation will be at a future TeV collider: 
for example, $e^- +e^+ \rightarrow t +\overline{c}$ at JLC:
$$
\begin{array}{ll}
\sigma =6.0\times 10^{-7}\ {\rm pb} & 
{\rm at}\ \sqrt{s}=0.2 \ {\rm TeV} \ , \\
\sigma =3.1\times 10^{-5}\ {\rm pb} & 
{\rm at}\ \sqrt{s}=2m_t=0.36 \ {\rm TeV} \ , \\
\sigma =1.1\times 10^{-4}\ {\rm pb} & 
{\rm at}\ \sqrt{s}=0.5 \ {\rm TeV} \ , \\
\sigma =7.5\times 10^{-4}\ {\rm pb} & 
{\rm at}\ \sqrt{s}=0.7 \ {\rm TeV} \ , \\
%\sigma =0.085\ {\rm pb} & 
%{\rm at}\ \sqrt{s}=m_{Z_R}=0.9 \ {\rm TeV} \ , \\
\end{array}
\eqno(6.2)
$$
where $\sigma=\sigma(t\overline{c})+\sigma(c\overline{t})$.

%%%%%%%%%%%%%%%%%%%%%%%%%%%%%%%%%%%%%% 7.
\vglue.2in

\noindent{\bf 7. Summary}
\vglue.05in

(i) Seesaw Mass Matrix  with 
$M_F$=[(unit matrix)+(rank-one matrix)]
can naturally understand the observed facts $m_t \gg m_b$ in 
contrast to $m_u \sim m_d$, and $m_t \sim \Lambda_W$.

(ii) Democratic seesaw mass matrix model  with the input $ b_e=0 $ 
can give reasonable quark mass ratios and CKM matrix 
by taking $b_u=-1/3$ and $b_d= - e^{i18^\circ}$, 
and a large neutrino mixing $\nu_\mu$-$\nu_\tau$ 
by taking $b_\nu \simeq -1/2$.

However, at present, we must take ad hoc parameter values
$b_u = -1/3$, $b_\nu \simeq -1/2$, $b_d \simeq -1$, $b_e = 0$.
I do not know whether there is some regularity among the values of $b_f$
or not, and what is the meaning of the parameter $b_f$.

(iii) The model will provide new physics in abundance:
\noindent
(a) $m_{t'}\sim$ a few TeV: we may expect a fourth up-quark production.
\noindent
(b) Abnormal structure of $U_R^u$: we may expect a single top-quark 
production.

However, whether these effects are visible or not in the near future 
depends on the value of $\kappa$  although we tentatively take $\kappa=10$ 
at the present study.
If $\kappa\simeq 10$, these effects cannot observe until starting of JLC.
Rather, there is a possibility that the effects due to the abnormal 
structure of $U_R^u$ are sensitive to the $K^0$-$\overline{K}^0$ 
mixing  which was pointed by T.~Kurimoto [18].
However, since our right-handed current structure 
is different from the conventional $SU(2)_L\times SU(2)_R$ models,
more careful study will be required.

(iv) Present model is still a semi-phenomenological model,
so that an embedding of the present model into a 
field-theoretical unification scenario is hoped.

%%%%%%%%%%%%%%%%%%%%%%%%%%%%%%%%%%%%%%%%%%%%%%%
\vglue.2in

\centerline{\large\bf Acknowledgments}

A series of works based on the democratic seesaw mass matrix model
was first started in collaboration with H.~Fusaoka.
The author would like to thank H.~Fusaoka for his enjoyable 
collaboration. 
This work was supported by the Grant-in-Aid for Scientific Research, the 
Ministry of Education, Science and Culture, Japan (No.08640386). 

%%%%%%%%%%%%%%%%%%%%%%%%%%%%%%%%%%%%%%%%%%%%%%%%%%%%%%%%%%%%%%%%%%%%%%%%%%%%
%%%%%%%%%%%%%%%%
\vglue.2in
\newcounter{0000}
\centerline{\large\bf References}
\begin{list}
{[~\arabic{0000}~]}{\usecounter{0000}
\labelwidth=0.8cm\labelsep=.1cm\setlength{\leftmargin=0.7cm}
{\rightmargin=.2cm}}
\item M.~Gell-Mann, P.~Rammond and R.~Slansky, in {\it Supergravity}, 
edited by P.~van Nieuwenhuizen and D.~Z.~Freedman (North-Holland, 
1979); 
T.~Yanagida, in {\it Proc.~Workshop of the Unified Theory and 
Baryon Number in the Universe}, edited by A.~Sawada and A.~Sugamoto 
(KEK, 1979); 
R.~Mohapatra and G.~Senjanovic, Phys.~Rev.~Lett.~{\bf 44}, 912 (1980).
\item Z.~G.~Berezhiani, Phys.~Lett.~{\bf 129B} (1983) 99;
Phys.~Lett.~{\bf 150B} (1985) 177;
D.~Chang and R.~N.~Mohapatra, Phys.~Rev.~Lett.~{\bf 58},1600 (1987); 
A.~Davidson and K.~C.~Wali, Phys.~Rev.~Lett.~{\bf 59}, 393 (1987);
S.~Rajpoot, Mod.~Phys.~Lett. {\bf A2}, 307 (1987); 
Phys.~Lett.~{\bf 191B}, 122 (1987); Phys.~Rev.~{\bf D36}, 1479 (1987);
K.~B.~Babu and R.~N.~Mohapatra, Phys.~Rev.~Lett.~{\bf 62}, 1079 (1989); 
Phys.~Rev. {\bf D41}, 1286 (1990); 
S.~Ranfone, Phys.~Rev.~{\bf D42}, 3819 (1990); 
A.~Davidson, S.~Ranfone and K.~C.~Wali, 
Phys.~Rev.~{\bf D41}, 208 (1990); 
I.~Sogami and T.~Shinohara, Prog.~Theor.~Phys.~{\bf 66}, 1031 (1991);
Phys.~Rev. {\bf D47} (1993) 2905; 
Z.~G.~Berezhiani and R.~Rattazzi, Phys.~Lett.~{\bf B279}, 124 (1992);
P.~Cho, Phys.~Rev. {\bf D48}, 5331 (1994); 
A.~Davidson, L.~Michel, M.~L,~Sage and  K.~C.~Wali, 
Phys.~Rev.~{\bf D49}, 1378 (1994); 
W.~A.~Ponce, A.~Zepeda and R.~G.~Lozano, 
Phys.~Rev.~{\bf D49}, 4954 (1994).
\item CDF Collaboration, F.~Abe {\it et al.}, Phys.~Rev.~Lett. 
{\bf 73}, 225 (1994).
\item Y.~Koide and H.~Fusaoka, Z.~Phys. {\bf C71} (1966) 459.
\item Y.~Koide and H.~Fusaoka, Prog.~Theor.~Phys. {\bf 97}, 459 (1966).
\item T.~Morozumi, T.~Satou, M.~N.~Rebelo and M.~Tanimoto, 
Preprint HUPD-9704 (1997), hep-ph/9703249.
\item T.~Satou, in this proceedings.
\item H.~Harari, H.~Haut and J.~Weyers, 
Phys.~Lett.~{\bf 78B}, 459 (1978);
T.~Goldman, in {\it Gauge Theories, Massive Neutrinos and 
Proton Decays}, edited by A.~Perlumutter (Plenum Press, New York, 
1981), p.111;
T.~Goldman and G.~J.~Stephenson,~Jr., Phys.~Rev.~{\bf D24}, 236 (1981); 
Y.~Koide, Phys.~Rev.~Lett. {\bf 47}, 1241 (1981); 
Phys.~Rev.~{\bf D28}, 252 (1983); {\bf 39}, 1391 (1989);
C.~Jarlskog, in {\it Proceedings of the International Symposium on 
Production and Decays of Heavy Hadrons}, Heidelberg, Germany, 1986
edited by K.~R.~Schubert and R. Waldi (DESY, Hamburg), 1986, p.331;
P.~Kaus, S.~Meshkov, Mod.~Phys.~Lett.~{\bf A3}, 1251 (1988); 
Phys.~Rev.~{\bf D42}, 1863 (1990);
L.~Lavoura, Phys.~Lett.~{\bf B228}, 245 (1989); 
M.~Tanimoto, Phys.~Rev.~{\bf D41}, 1586 (1990);
H.~Fritzsch and J.~Plankl, Phys.~Lett.~{\bf B237}, 451 (1990); 
Y.~Nambu, in {\it Proceedings of the International Workshop on 
Electroweak Symmetry Breaking}, Hiroshima, Japan, (World 
Scientific, Singapore, 1992), p.1.
\item N.~Cabibbo, Phys.~Rev.~Lett.~{\bf 10}, 531 (1996); 
M.~Kobayashi and T.~Maskawa, Prog.~Theor.~Phys.~{\bf 49}, 652 (1973).
\item Y.~Koide,  Mod.~Phys.~Lett.{\bf A8}, 2071 (1993).
\item Y.~Koide, Lett.~al Nuovo Cim. {\bf 34}, 201 (1982); Phys.~Lett. 
{\bf 120B}, 161 (1983); Phys.~Rev. {\bf D28}, 252 (1983).
\item Y.~Koide,  Mod.~Phys.~Lett.{\bf A5}, 2319 (1990); 
Y.~Koide and M.~Tanimoto, Z.~Phys. {\bf C72}, 333 (1996).
\item Particle data group, R.~M.~Barnet {\it et al.}, Phys.~Rev. 
{\bf D54}, 1 (1996).
\item Y.~Fukuda {\it et al.}, Phys.~Lett. {\bf B335}, 237 (1994).
\item For example, P.~Anselmann {\it et al.}, GALLEX collaboration, 
Phys.~Lett. {\bf B327}, 377 (1994); {\bf B357}, 237 (1995); 
J.~N.~Abdurashitov {\it et al.}, SAGE collaboration, {\it ibid.} 
{\bf B328}, 234 (1994). See also, N.~Hata and P.~Langacker, 
Phys.~Rev. {\bf D50}, 632 (1994); {\bf D52}, 420 (1995).
\item Y.~Koide,  Mod.~Phys.~Lett.{\bf A36}, 2849 (1996).
\item Y.~Koide, Preprint US-96-09 (1996), hep-ph/9701261, unpublished.
\item T.~Kurimoto, private communication. Also see, T.~Kurimoto, 
A.~Tomita and S.~Wakaizumi, Phys.~Lett. {\bf B381}, 470 (1996).
%\item J.~C.~Pati and A.~Salam, Phys.~Rev. {\bf D10}, 275 (1974); 
%R.~N.~Mohapatra and J.~C.~Pati, Phys.~Rev. {\bf D11},
% 366 and 2588 (1975);
%G.~Senjanovic and R.~N.~Mohapatra, Phys.~Rev. {\bf D12}, 1502 (1975).
%
%\item Y.~Kimura, presented at The 5th ICFA Seminar at KEK {\it ``Future 
%Perspectives in High Energy Physics"}, Oct. 15-18, 1996, to be published in 
%the Proceedings.
%\item H.~Terazawa, University of Tokyo, Report No.~INS-Rep.-298 (1977) 
%(unpublished); Genshikaku Kenkyu (INS, Univ.~of Tokyo) {\bf 26}, 33 
%(1982);
%Y.~Koide, Phys.~Rev. {\bf D49}, 2638 (1994).
%
\end{list}
%%%%%%%%%%%%%%%

\end{document}